 \documentclass[final,3p,times,twocolumn]{elsarticle}

\usepackage{pgfplots}
\usepackage{pgf}
\usepackage{tikz}
\usepackage{graphicx}
\usepackage[]{subfigure}
\usepackage{lineno}

\begin{document}

\title{\textit{Ab Initio} Study of the Magnetic Behavior of Metal Hydrides: A Comparison with the Slater-Pauling Curve}

\author{Andrea Le\'on$^{1,4}$}
\author{E. A. Vel\'asquez$^{2,3,4}$}
\author{J. Mej\'ia-L\'opez$^{2,4}$}
\author{P. Vargas$^{1,4}$}

\address{$^1$Departamento de F\'isica, Universidad T\'ecnica Federico Santa Mar\'ia, Valpara\'iso, Chile\\ 
$^2$Centro de Investigaci\'on en Nanotecnolog\'ia y Materiales Avanzados CIEN-UC, Facultad de F\'isica, 
Pontificia Universidad \\ Cat\'olica de Chile, Santiago, Chile\\ 
$^3$ Grupo de Investigaci\'on en Modelamiento y Simulaci\'on Computacional, 
Universidad de San Buenaventura \\ Seccional Medell\'in, Medell\'in, Colombia\\
$^4$Centro para el Desarrollo de la Nanociencia y la Nanotecnolog\'ia CEDENNA, Santiago, Chile}


\begin{abstract}
We investigated the magnetic behavior of metal hydrides FeH$_{x}$, CoH$_{x}$ and NiH$_{x}$ for several concentrations of 
hydrogen ($x$) by using Density Functional Theory calculations. Several structural phases of the metallic host: 
bcc ($\alpha$), fcc ($\gamma$), hcp ($\varepsilon$), dhcp ($\varepsilon'$),  tetragonal structure for FeH$_{x}$ and 
  $\varepsilon$-$\gamma$ phases for CoH$_{x}$, were studied. We found that for CoH$_{x}$  and NiH$_{x}$ the magnetic moment 
($m$) decreases regardless the concentration $x$. However, for FeH$_{x}$ systems, $m$ increases or decreases depending on
the variation in $x$. In order to find a general trend for these changes of $m$ in magnetic metal hydrides, we compare our 
results with the Slater-Pauling curve for ferromagnetic metallic binary alloys. It is found  that the $m$ of metal hydrides made 
of Fe, Co and Ni fits the shape of the Slater-Pauling curve as a function of $x$. Our results indicate that there are two main 
effects that determine the $m$ value due to hydrogenation: an increase of volume causes $m$ to increase, and the 
addition of an extra electron to the metal always causes it to decrease. We discuss these behaviors in detail.
\end{abstract}


\maketitle
\section{Introduction}

Magnetic hydrides (MH$_x$) based on 3d-metal alloys have been intensively studied for decades. They are of particular 
interest due to the different variations of the magnetic moment ($m$) of Fe, Ni and Co hydrides and their alloys as 
functions of the hydrogen concentration ($x$). There is a complete experimental study by 
Antonov \textit{et al.} \cite{Co1,1Exp,2Exp,Co2,Abook} describing in detail the magnetic properties and magnetic phase 
diagram at 7-8 GPa and temperatures around 700 K for Fe and Co hydrides, with an H:metal ratio close to 1 \cite{Co1,Abook}.

Moreover, the FeH$_{x}$ systems are of particular interest from a geophysical point of view for processes in the Earth's 
inner core. Related to this issue, recent works have reported synthesis of $\gamma$-FeH$_{x}$ at high pressures and 
temperatures with stoichiometric composition and low concentrations ($x$ $<$ 1) \cite{Olga,3Exp}. Some reports explored the 
formation of new structures at higher concentrations ($x$ $>$ 1) as well \cite{4Exp}. Thus, new phase diagrams have been 
established at high temperatures from 0 to 1200 K and high pressures from 0-10 GPa \cite{3Exp} and 0-120 GPa  \cite{4Exp}. 
Under these conditions, new structures based on FeH$_{x}$ ($x$ $>$ 1) that exhibit magnetic and non-magnetic orderings have been found \cite{4Exp}. 

On the other hand, the case of CoH$_{x}$ is of particular interest due to the study of single molecular magnets and anisotropy 
in magnetic molecular junctions for a high concentration ($x$ = 1,2) \cite{SCH2,SCH3}. 

Hence, the study of the magnetic moments of FeH$_{x}$ and CoH$_{x}$ as a function of the H concentration in ferromagnetic 
metals is a relevant issue. Indeed, several electronic structure calculations for FeH and CoH hydrides have been carried 
out previously, where the structural stabilities and magnetic properties have been studied \cite{elsse,DFTCo1}. Concerning 
the limit of low concentration one can find studies on samples of FeH$_x$ for $x$ = 0.03 and $x$ = 0.25, 0.5, 0.75  \cite{FeHST,conf}. 
However, First-Principles-based studies on the magnetic moment in a wide range of concentrations, for both the previously-known and 
new phases of FeH$_{x}$ \cite{4Exp}, have yet to be reported.

The magnetic properties of materials are strongly affected by their hydrogen absorption, normally leading to a reduction in 
the $m$ of intermetallic compounds based on Co and Ni but to an increase in the $m$ of Fe. Indeed, experimental works 
on these metals and their alloys    \cite{Co1,1Exp,2Exp,Co2,Abook} and Density Functional Theory (DFT) studies show 
these behaviors \cite{elsse,DFTCo1,FeHST,conf,stoner}. This evident $x$ dependence on the magnetic properties of metal 
hydrides is accompanied by the increase of volume produced by the inclusion of H atoms within the metal. When $x$ 
increases, the volume also increases, which generates a localization of the electronic states, and, an increase 
in the value of $m$ is therefore expected. Additionally, the charge transference between the metal and the hydrogen can decrease or increase 
the magnetic moment, $m$, of the metal atom depending on whether or not the minority or majority spin band is full.

In order to study these effects, that modify the magnetic properties, we performed spin-dependent DFT calculations on 
the Fe and Co hydrides'systems. We considered different structural phases such as $\alpha$, $\gamma$, $\varepsilon$, 
$\varepsilon'$ and tetragonal with a space group of $I4/mmm$ for FeH$_x$ systems, and $\varepsilon$ and $\gamma$ for the 
CoH$_x$ systems. 
The antiferromagnetic (AFM) phase of the $\varepsilon$-FeH$_{x}$ and $\gamma$-FeH$_{x}$ structures have also been studied  
in order to comprehend the stability between the AFM and ferromagnetic (FM) phases as a function of $x$. 
Additionally, the calculations of the new phase with the space group $Pm$-$\bar{3}$$m$ at 
a high H concentration ($x$ = 3) in Fe, reported in Ref. \cite{4Exp}, have been reported in this study. 

For comparison reasons, we superimpose our results on the Slater-Pauling (SP) curve \cite{SP}. The SP curve shows the magnetic 
moment per atom of an alloy as a function of the number of valence electrons. It was constructed for binary metallic alloys, 
and the approach has also been extended to model the magnetic moment of a magnetic metal host with non-metallic interstitial 
impurities \cite{matar1,matar2}. In this work, we used the SP curve for the MH$_x$ system within the framework of the d-rigid 
band model \cite{1Exp,Abook}. 

\section{COMPUTATIONAL DETAILS}
 In this study, the Vienna \textit{Ab Initio} Simulation Package (VASP) \cite{Vasp} is used, with the Generalized Gradient 
 Approximation (GGA) along with the Perdew-Burke-Ernzerhof (PBE) \cite{PBE} prescription. As a result of the convergence tests, 
 we used a kinetic energy cutoff of 360 eV for all the calculations, and Brillouin k-points grids of 8$\times$8$\times$8 in 
 $\alpha$, 8$\times$8$\times$4 in $\varepsilon$, 6$\times$6$\times$6 in $\gamma$ and 12$\times$12$\times$6 in 
 both the $\varepsilon'$ and tetragonal structures. The calculations were performed by considering spin-polarized configurations, 
 FM and AFM, in order to obtain the magnetic configuration with the lowest energy.  The  relaxed  structures were obtained  
 when forces on each atom were less than 0.01 eV/\AA$ $.
 
\section{CRYSTAL STRUCTURES}

For the $\alpha$ and $\varepsilon$ phases, we used a super cell of 16 Fe atoms, including 16 interstitial sites with 
tetrahedral ($\alpha$) and octahedral ($\varepsilon$) symmetry, which we filled with H in different concentrations.  
In the $\gamma$ and $\varepsilon'$ structures of Fe, we used a super cell with 32 and 4 atoms with 32 and 4 interstitial sites 
with octahedral symmetry respectively. For Co,  we used  $\varepsilon$ and $\gamma$ super cell structures with 16 and 
4 Co atoms, with 16 and 4 interstitial sites with octahedral symmetry, respectively. 

The starting aspect ratios were  $c/a$ = $\sqrt{8/3}$ and $c/a$ = 2$\times$$\sqrt{8/3}$ for the $\varepsilon$ and 
the $\varepsilon'$ structures, respectively, while we used the unit cell reported in Ref. \cite{4Exp} for the tetragonal phase of Fe.
The unit cells for the obtained crystal structures are shown in Fig \ref{FIG_1}.

\begin{figure}[h]
\centering
\subfigure[bcc ($\alpha$)]{\includegraphics[width=1.3in]{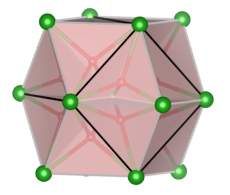}}\subfigure[fcc ($\gamma$)]{\includegraphics[width=1.3in]{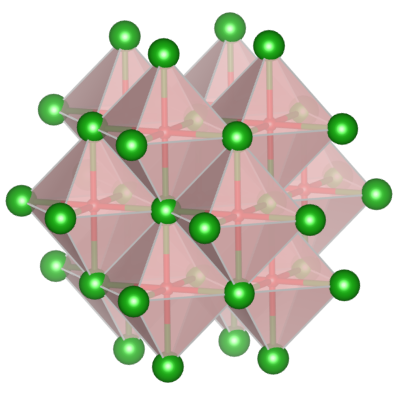}}
\label{FIG_1}
\end{figure}
\vspace{-0.8cm}
\begin{figure}[h]
\subfigure[hcp ($\varepsilon$)]{\includegraphics[width=1.1in]{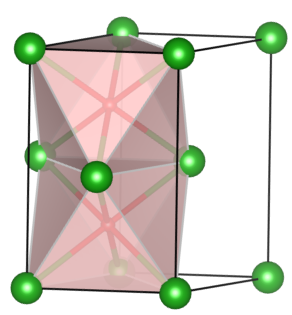}}
\subfigure[dhcp ($\varepsilon'$)]{\includegraphics[width=0.95in]{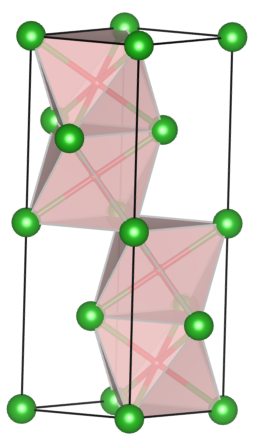}}\subfigure[$I4/mmm$]{\includegraphics[width=0.98in]{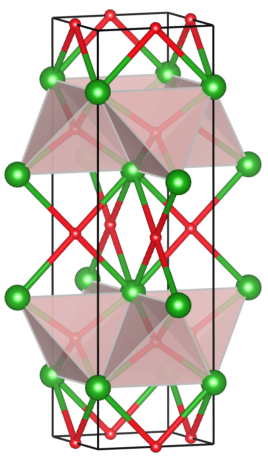}}
\caption{The  different calculated crystal structures. Green spheres depict metal atoms coordinated with the H atoms filling  the interstitial sites with tetrahedral and octahedral symmetry for the $\alpha$, $\gamma$, $\varepsilon$, $\varepsilon'$ and tetragonal structures, as applicable. In the tetragonal structure H atoms fill sites with tetrahedral symmetry, but some H atoms are also connected with four metal atoms in the same plane.}
\label{FIG_1}
\end{figure}

\section{Results and Discussions}
\begin{figure}[h]
 \centering
 \definecolor{color1}{RGB}{30,144,255}
\definecolor{gris0}{RGB}{240,240,240}
\definecolor{gris1}{RGB}{230,230,230}
\definecolor{gris2}{RGB}{200,200,200}
\definecolor{gris3}{RGB}{112,128,144}
\definecolor{gris4}{RGB}{120,120,120}
\definecolor{gris5}{RGB}{255,255,240}

\definecolor{hcp}{RGB}{255,255, 0}
\definecolor{fcc}{RGB}{112,138,144}
\definecolor{I4}{RGB}{255,127, 80}
\definecolor{I42}{RGB}{50,205, 50}
\definecolor{cohcp}{RGB}{0,255,0}
\definecolor{ni}{RGB}{255, 20,147}
\definecolor{co1}{RGB}{199, 21,133}
\definecolor{fcc2}{RGB}{0,0,139}
\definecolor{fcc3}{RGB}{255,160,122}
\definecolor{fcc4}{RGB}{72,209,204}
\definecolor{gg}{RGB}{230,230, 230}
\definecolor{indi}{RGB}{75,0,130}
\definecolor{rp}{RGB}{0,255,255}
\definecolor{naranja}{RGB}{255, 69,0}

	\begin{tikzpicture}
		\begin{axis}[
		 grid=major,
		 grid style = {color=gg},
			width=0.95\columnwidth,
			ymin=-0.1,
			ymax=2.5,
			xmin=-0.1,
			xmax=2.1,
			ytick distance=0.5,
			xtick distance=0.5,
            every axis y label/.style={at={(ticklabel cs: 0.5, 5.5)},rotate=90,anchor=center},
		xlabel=$\textbf{\textit{x}}$,
		ylabel=\textbf{$m\hspace{0.08cm} (\mu_{B})$},
			legend style={font=\scriptsize,draw=black},
			legend entries={$\gamma$-NiH$_{x}$,$\varepsilon$-CoH$_{x}$,$\gamma$-CoH$_{x}$, $\alpha$-FeH$_{x}$, $\varepsilon$-FeH$_{x}$,$\gamma$-FeH$_{x}$,$\varepsilon'$-FeH$_{x}$,$I4/mmm$-FeH$_{x}$, $Pm$-$\bar{3}$$m$-FeH$_{3}$},
			legend columns=1,
			legend to name=named
		]
		\addplot[mark=square*,only marks,mark size=1.8pt,mark options={fill=ni,draw=black,rotate=45}] table[x=x,y=m] {magNi.txt};
		\addplot[mark=square*,only marks,mark size=2.0pt,mark options={fill=cohcp,draw=black,rotate=90}] table[x=x,y=m] {magCohcp.txt};
		\addplot[mark=triangle*,only marks,mark size=2.5pt,mark options={fill=indi,draw=black,rotate=90}] table[x=x,y=m] {magCofcc.txt};
		\addplot[gris2,mark=*,only marks,mark size=2.2pt,mark options={fill=color1,draw=black}] table[x=x,y=m] {magbcc.txt};
		\addplot[mark=square*,only marks,mark size=2.3pt,mark options={fill=hcp,draw=black}] table[x=x,y=m] {maghcp.txt};
		\addplot[mark=triangle*,only marks,mark size=2.5pt,mark options={fill=I42,draw=black}] table[x=x,y=m] {magfcc.txt};
		\addplot[mark=triangle*,only marks,mark size=2.4pt,mark options={fill=red,draw=black,rotate=-180}] table[x=x,y=m] {magdhcp.txt};
		\addplot[mark=triangle*, only marks,mark size=2.6pt,mark options={fill=fcc4,draw=black,rotate=-90}] table[x=x,y=m] {magI4.txt};
		\addplot[mark=pentagon*,only marks,mark size=2.8pt,mark options={fill=rp,draw=black}] table[x=x,y=m] {magPm3m.txt};			
		\node at (axis cs: 1.8, 1.1) {(a)};
         	\end{axis}
	\end{tikzpicture}
	
\vspace{-2.5cm}
\hspace{3.8cm}
	\begin{tikzpicture}
		\begin{axis}[
		 grid style = {color=gg},
                 width=0.40\columnwidth, height=0.30\columnwidth,
		ymin=-0.1,
		ymax=0.1,
		xmin=2.5,
		xmax=3.5,
		ytick distance=0.1,
		xtick distance=0.5,
		x tick label style={font=\tiny},
		y tick label style={font=\tiny},		
		]
		\addplot[mark=pentagon*,only marks,mark size=2.8pt,mark options={fill=rp,draw=black}] table[x=x,y=m] {magPm3m.txt};
		\end{axis}
		\end{tikzpicture}
		
\vspace{1.2cm}		
\hspace{-0.28cm}
	\begin{tikzpicture}
		\begin{axis}[
		grid=major,
		 grid style = {color=gg},
                 width=0.7\columnwidth, height=0.6435\columnwidth,
		ymin=-0.1,
		ymax=2.5,
		xmin=-0.1,
		ytick distance=0.5,
		xtick distance=0.25,
		xlabel=$\textbf{\textit{x}}$,
		ylabel=\textbf{$m\hspace{0.08cm} (\mu_{B})$},
		x tick label style={font=\scriptsize},
		y tick label style={font=\scriptsize},
		every axis y label/.style={at={(ticklabel cs: 0.5, 5.5)},rotate=90,anchor=center},
		]
		\addplot[mark=square*,only marks,mark size=1.5pt,mark options={fill=ni,draw=black,rotate=45}] table[x=x,y=m] {magNi.txt};
		\addplot[mark=square*,only marks,mark size=1.5pt,mark options={fill=white,draw=black,rotate=45}] table[x=x,y=m] {expNi.txt};
		\addplot[mark=halfsquare*,only marks,mark size=2.5pt,mark options={fill=white,draw=black,rotate=45}] table[x=x,y=m] {ECo_hcp.txt};
		\addplot[mark=square*,only marks,mark size=2.0pt,mark options={fill=cohcp,draw=black,rotate=90}] table[x=x,y=m] {dftCohcp.txt};
		\addplot[gris2,mark=triangle,only marks,mark size=2.5pt,mark options={fill=black,draw=black,rotate=90}] table[x=x,y=m] {ECo_fcc.txt};
		\addplot[mark=triangle*,only marks,mark size=2.5pt,mark options={fill=indi,draw=black,rotate=90}] table[x=x,y=m] {dftCohfcc.txt};
		\addplot[mark=triangle*,only marks,mark size=2.5pt,mark options={fill=white,draw=black,rotate=-180}] table[x=x,y=m] {dftFedhcp.txt};
		\addplot[mark=triangle*,only marks,mark size=2.5pt,mark options={fill=red,draw=black,rotate=-180}] table[x=x,y=m] {Efe_dhcp.txt};
		\addplot[mark=square*, only marks,mark size=2.5pt,mark options={fill=hcp,draw=black,rotate=-90}] table[x=x,y=m] {dftFehcp.txt};
		\addplot[mark=square*,only marks,mark size=1.8pt,mark options={fill=white,draw=black}] table[x=x,y=m] {Efe_hcp.txt};
		\addplot[mark=*,only marks,mark size=2.0pt,mark options={fill=white,draw=black}] table[x=x,y=m] {Efe.txt};
		\addplot[mark=*,only marks,mark size=1.8pt,mark options={fill=color1,draw=black}] table[x=x,y=m] {dftfe.txt};			
		\node at (axis cs: 0.9,0.35) {(b)};
		\end{axis}
		\node at (5.1,1.68) {\ref{named}};
		\end{tikzpicture}
 \caption{(a) Magnetic moment per metal atom (in $ \mu _B$ units) as a function of the concentration of 
  $x$ in FeH$_x$, for $\alpha$, $\gamma$, $\varepsilon$, $\varepsilon'$  and tetragonal structure ($I4/mmm$), 
  and $\gamma$ and $\varepsilon$ phases for CoH$_{x}$. Simple cubic ($Pm$-$\bar{3}$$m$) FeH$_{3}$  
  and $\gamma$-NiH$_x$ reported by \cite{4Exp,stoner} are included (cyan pentagon and pink diamonds, 
  respectively). (b) Comparison between our results and experimental data (empty symbols for each structure) as reported in \cite{Co1,1Exp,2Exp,Co2,Abook}. The equilibrium volume as a function of $x$ for the MH$_{x}$ systems is given in the Figs. 1-3 in \cite{dataB}.}
\label{fig2}
\end{figure}
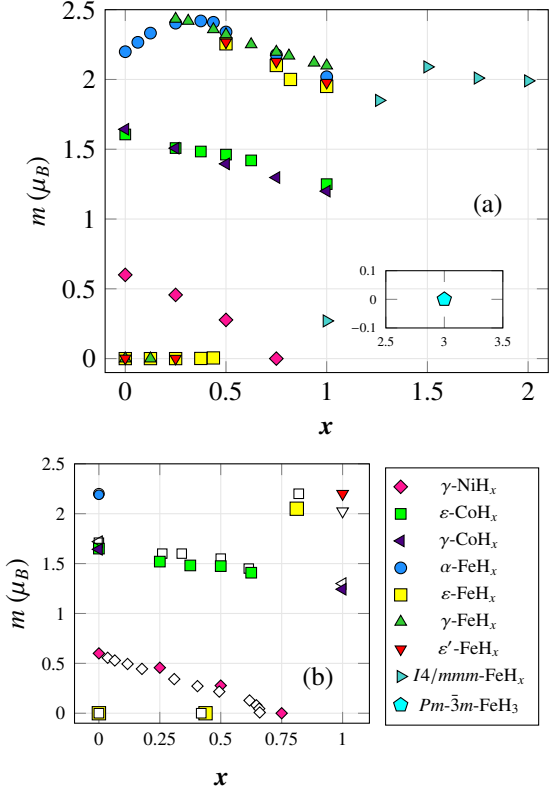

Figure \ref{fig2} (a) shows $m$ as a function of the concentration, $x$, for the samples FeH$_{x}$ in $\alpha$, $\gamma$, 
$\varepsilon$, $\varepsilon'$, tetragonal and simple cubic phases. The $\gamma$ and $\varepsilon$ phases for 
CoH$_{x}$ are reported as well. We also include  the recently-reported results of NiH$_x$ in the $\gamma$ phase by 
using the same reported technique \cite{stoner}. Additionally, in Fig. \ref{fig2} (b), we show the comparison with experimental 
data obtained at some concentrations (empty  symbols \cite{Co1,1Exp,2Exp,Co2,Abook}). In general, our results and the experimental ones
match quite well. Further, we report the energy values and compare the structural stabilities. 

As observed,  for FeH$_{x}$  at $x$ $<$ 0.75, the  
$\alpha$-phase presents the lowest energy. However, for $x$=1, the $\varepsilon'$-phase has  lower energy; following 
in order of energy are: E$_{\varepsilon'}$$ < $E$_{\varepsilon}$$ < $E$_{\gamma}$$ < $E$_\alpha$.  
For CoH$_{x}$, the $\varepsilon$-phase is the lowest energy structure when $x$ $\leq$ 0.5, whereas the 
$\gamma$-phase presents the lowest energy when the concentration goes beyond 0.75 ($x$ $\geq$ 0.75). 

It is worth noting that the FM phase is the most stable magnetic configuration for several systems. 
However, for the $\varepsilon$-FeH$_{x}$ structure, we found that for $x$ $\leq$ 0.44, the AFM phases are the most stable.  The $\varepsilon'$ structure exhibits 
paramagnetic behavior for $x$ $\leq$ 0.25 and FM for $x$ $\geq$ 0.5. Similar behavior is observed in the 
$\gamma$-FeH$_{x}$ structure, which presents an AFM phase for $x$ = 0 and an FM phase when $x$  $\geq$ 0.25.\\

The calculated magnetic systems shown in Fig. 2a display two different remarkable behaviors. On one hand,
there is always a steady decrease of $m$ when $x$ increases in systems based on Co and Ni. 
On the other hand, for the $\alpha$-FeH$_x$ phase, the $m$ value increases for 
$0 \leq$ $x$ $\leq$ 0.375 towards a maximum, and, $m$ then decreases, as it does in the cases of Co and Ni. 
In the cases of $\gamma$-, $\varepsilon$- and $\varepsilon'$-FeH$_x$ with $x$ $>$ 0.25 ($\gamma$) and $x$ $>$ 0.5 ($\varepsilon$, $\varepsilon'$ ) respectively, the magnetic moment always decreases. For $I4/mmm$-FeH$_x$ with $x$ $>$ 1,  the magnetic moment increases in the range of 1 $\leq$ $x$ $\leq$ 1.5 and decreases for 1.5 $<$ $x$ $\leq$ 2. Fitted curves of the magnetic moment for FeH$_x$, CoH$_x$ and NiH$_x$ as a function of $x$, are given in the Figs. 4-5 in \cite{dataB}.

\begin{figure}
 \centering
  \definecolor{color1}{RGB}{30,144,255}
\definecolor{gris0}{RGB}{240,240,240}
\definecolor{gris1}{RGB}{230,230,230}
\definecolor{gris2}{RGB}{200,200,200}
\definecolor{gris3}{RGB}{112,128,144}
\definecolor{gris4}{RGB}{120,120,120}
\definecolor{gris5}{RGB}{255,255,240}

\definecolor{hcp}{RGB}{255,255, 0}
\definecolor{fcc}{RGB}{112,138,144}
\definecolor{I4}{RGB}{0,255,255}
\definecolor{cohcp}{RGB}{0,255,0}
\definecolor{ni}{RGB}{255, 20,147}
\definecolor{co1}{RGB}{199, 21,133}
\definecolor{fcc2}{RGB}{0,0,139}
\definecolor{gg}{RGB}{230,230, 230}
\definecolor{I42}{RGB}{50,205, 50}

	\begin{tikzpicture}
		\begin{axis}[
		 grid=major,
		 grid style = {color=gg},
			width=0.9\columnwidth,
			ymin= -0.1,
			ymax=2.75,
			xmin=-0.05,
			xmax=1.05,
			ytick distance=0.5,
			xtick distance=0.25,
            every axis y label/.style={at={(ticklabel cs: 0.5, 5.5)},rotate=90,anchor=center},
			xlabel=$\textbf{\textit{x}}$,
			ylabel=\textbf{$m \hspace{0.08cm}(\mu_{B})$},
			legend style={font=\tiny,draw=black},
		]
		\addplot[dashed,black,mark=halfcircle*, mark size=2.9pt, mark options={fill=red,draw=black, solid}] table[x=x,y=m] {m_Fe_vol_FeHx.txt};
		\addplot[dashed,black,mark=halfsquare left*, mark size=3.0pt, mark options={solid,fill=I42,draw=black,rotate=45}] table[x=x,y=m] {m_Co_vol_CoHx.txt};
		\addplot[dashed,black,mark=*, mark size=2.8pt, mark options={solid,fill=color1,draw=black}] table[x=x,y=m] {magbcc.txt};
		\addplot[dashed,black,mark=square*,mark size=2.2pt,mark options={solid,fill=cohcp,draw=black,rotate=90}] table[x=x,y=m] {magCohcp.txt};
                 \addplot[dashed,black,mark=*,mark size=2.8pt,mark options={solid,fill=white,draw=black}] table[x=x,y=m] {m_FeH_vol_fijo.txt};
		\addplot[dashed,mark=halfsquare left*,mark size=3.0pt,mark options={solid,fill=white,draw=black,rotate=45}] table[x=x,y=m] {m_CoH_vol_fijo.txt};
		\addplot[dashed,mark=halfsquare left*,mark size=3.0pt,mark options={solid,fill=white,draw=black,rotate=90}] table[x=x,y=m] {m_NiH_vol_fijo.txt};
                \addplot[dashed,black,mark=halfsquare left*, mark size=3.0pt, mark options={solid,fill=ni,draw=black,rotate=90}] table[x=x,y=m] {m_Ni_vol_NiHx.txt};
                \addplot[dashed,black,mark=square*,mark size=2.2pt,mark options={solid,fill=ni,draw=black,rotate=45}] table[x=x,y=m] {magNi.txt};	
	        \node at (axis cs: 0.5,1) {};
         	\end{axis}
	\end{tikzpicture}

 \caption{Magnetic moment (in $ \mu _B$ units) per metal atom as a function on $x$ for $\alpha$-FeH$_{x}$ (circles), 
 $\varepsilon$-CoH$_{x}$ (squares), and $\gamma$-NiH$_x$ (diamonds). Filled symbols 
 represent the magnetic moment of MH$_x$ systems at the equilibrium volume, as shown in Fig. \ref{fig2}. 
 Semi-filled symbols depict the magnetic moment of pure metals (no hydrogen) calculated at the equilibrium 
 of the MH$_x$ system. Empty symbols show the magnetic moment for MH$_x$ systems computed at the equilibrium 
 volume of the pure metal.}
 \label{mag}
\end{figure}
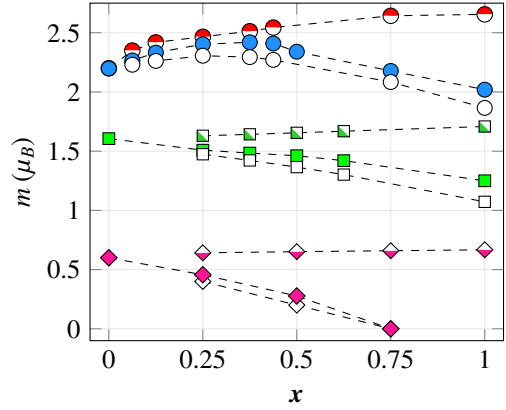

As a general feature, when H atoms fill the interstitial places of a metal lattice, new electronic states are created at low 
energy with respect to the Fermi level, giving rise to a reordering of the electronic states of the metal, thus affecting the 
magnetic properties. Another way to see this is to think of a Stoner-type of model \cite{stoner}, where the hybridization 
between the metal and H atoms changes the band energy (kinetic energy) and affects the magnetic energy, thus leading 
to a reduction or increase of the magnetic moment. 

Besides, it is already known that H atoms inserted in the interstitial places of a metallic host, produce a volume increase. 
The volume expansion has the effect of localizing the electronic states, thus narrowing the bandwidth for both spins, which 
generally favors the stability of the ferromagnetic phase.  To quantify the effect of the volume expansion, 
we performed our \textit{Ab Initio} 
calculations for pure metal systems incorporating the volume expansion produced by the H atoms. These results are 
shown in Fig. \ref{mag}  with semi-filled symbols. A steady increase for $m$ is clearly observed for the pure metal 
systems, indicating the effect of volume change on $m$. In all cases, the expansion of metal produces a bigger $m$ 
than that found in MH$_x$ systems. 

Moreover, if the volume in the MH$_x$ systems is kept fixed to the value of the pure metal (empty symbols in Fig. \ref{mag}), 
the $m$ always diminishes due to the presence of H. These results indicate that there are two competing effects that determine 
the value of $m$: solely increasing the volume always causes an increase of $m$, but adding an extra electron to the metal 
always causes a decrease of $m$.

\begin{figure}
 \centering
 \definecolor{color1}{RGB}{30,144,255}
\definecolor{gris0}{RGB}{240,240,240}
\definecolor{gris1}{RGB}{230,230,230}
\definecolor{gris2}{RGB}{200,200,200}
\definecolor{gris3}{RGB}{112,128,144}
\definecolor{gris4}{RGB}{120,120,120}
\definecolor{gris5}{RGB}{255,255,240}

\definecolor{hcp}{RGB}{255,255, 0}
\definecolor{fcc}{RGB}{112,138,144}
\definecolor{I4}{RGB}{255,127, 80}
\definecolor{cohcp}{RGB}{0,255,0}
\definecolor{ni}{RGB}{255, 20,147}
\definecolor{co1}{RGB}{199, 21,133}
\definecolor{fcc2}{RGB}{0,0,139}
\definecolor{rp}{RGB}{0,255,255}
\definecolor{fcc4}{RGB}{72,209,204}
\definecolor{I42}{RGB}{50,205, 50}
\definecolor{gg}{RGB}{230,230, 230}
\definecolor{indi}{RGB}{75,0,130}

\begin{tikzpicture}
		\begin{axis}[
		minor tick num=3,
		grid=major,
		grid style = {color=gg},
		width=0.97\columnwidth,
		ymin=-0.1,
		ymax=2.6,
         every axis y label/.style={at={(ticklabel cs: 0.5, 5.5)},rotate=90,anchor=center},
		ytick distance=0.5,
		xlabel=\textbf{$N$},
		ylabel=\textbf{$m \hspace{0.08cm}  (\mu_{B})$},
		legend style={font=\scriptsize,draw=black},
		legend entries={CoCr, CoMn, FeCo, FeCr, FeNi, FeV, NiCo, NiCr, NiCu, NiMn, NiV, NiZn},
		legend columns=4,
		legend to name=named
		]
		\addplot[gris1,mark=halfsquare left*,mark size=2pt,mark options={fill=gris3,draw=gris2}] table[x=N,y=m] {CoCr.txt};
		\addplot[gris1,mark=10-pointed star,mark size=2pt,mark options={fill=gris4,draw=gris2}] table[x=N,y=m] 
{CoMn.txt};
	\addplot[gris1,mark=*,mark size=2pt,mark options={fill=gris0,draw=gris2}] table[x=N,y=m] 
{FeCo.txt};
	\addplot[gris2,mark=halfsquare*,mark size=2pt,mark options={fill=gris3,draw=gris2}] table[x=N,y=m] 
{FeCr.txt};
\addplot[gris1,mark=diamond,mark size=2pt,mark options={fill=gris1,draw=gris2}] table[x=N,y=m] 
{FeNi.txt};
\addplot[gris2,mark=x,mark size=2pt,mark options={fill=gris4,draw=gris4}] table[x=N,y=m] 
{FeV.txt};
\addplot[gris2,mark=square,mark size=2pt,mark options={fill=gris1,draw=gris2}] table[x=N,y=m] 
{NiCo.txt};
\addplot[gris2,mark=diamond,mark size=2pt,mark options={fill=gris1,draw=gris2}] table[x=N,y=m] 
{NiCr.txt};
\addplot[gris2,mark=triangle,mark size=2pt,mark options={fill=gris1,draw=gris2}] table[x=N,y=m] 
{NiCu.txt};
\addplot[gris2,mark=pentagon,mark size=2pt,mark options={fill=gris2,draw=gris3}] table[x=N,y=m] 
{NiMn.txt};
\addplot[gris2,mark=halfdiamond*,mark size=2pt,mark options={fill=gris1,draw=gris2}] table[x=N,y=m] 
{NiV.txt};
\addplot[gris2,mark=halfsquare right*,mark size=2pt,mark options={fill=gris1,draw=gris2}] table[x=N,y=m] 
{NiZn.txt};
\addplot[gris2,mark=*,only marks,mark size=2.5pt,mark options={fill=color1,draw=black}] table[x=N,y=m] {sp1_bcc.txt};
		\addplot[mark=triangle*,only marks,mark size=2.9pt,mark options={fill=fcc4,draw=black}] table[x=N,y=m] {sp2_fcc.txt};
		\addplot[mark=square*,only marks,mark size=2.2pt,mark options={fill=hcp,draw=black}] table[x=N,y=m] {sp3_hcp.txt};
		\addplot[mark=triangle*,only marks,mark size=2.8pt,mark options={fill=red,draw=black,rotate=-180}] table[x=N,y=m] {sp4_dhcp.txt};
		\addplot[mark=triangle*, only marks,mark size=2.8pt,mark options={fill=I42,draw=black,rotate=-90}] table[x=N,y=m] {sp5_I4.txt};
		\addplot[mark=square*,only marks,mark size=2.0pt,mark options={fill=cohcp,draw=black,rotate=90}] table[x=N,y=m] {sp7_hcp_co.txt};
		\addplot[mark=triangle*,only marks,mark size=2.5pt,mark options={fill=indi,draw=black,rotate=90}]  table[x=N,y=m] {sp8_fcc_co.txt};
		\addplot[mark=square*,only marks,mark size=1.8pt,mark options={fill=ni,draw=black,rotate=45}] table[x=N,y=m] {sp9_fcc_ni.txt};
        \addplot[mark=pentagon*,only marks,mark size=2.8pt,mark options={fill=rp,draw=black}] table[x=N,y=m] {sp6_Pm3m.txt};
		\end{axis}
        \node at (3,5.8) {\ref{named}};
		\end{tikzpicture}
 \caption{Magnetic moment as a function of the effective valence electrons ($N$) for MH$_{x}$. The same colors and symbols 
 for the systems shown in Fig. \ref{fig2} are used. Data are superimposed on the Slater-Pauling curve (grey symbols) 
 digitized from \cite{SP}.  $N = N_{M}+\eta x$, where N$_{M}$ = 8(Fe), 9(Co) and 10(Ni), $x$ is the hydrogen 
 concentration, and $\eta$ is the effective charge per H atom transferred to the metal.
 }
\label{fig3}
\end{figure}
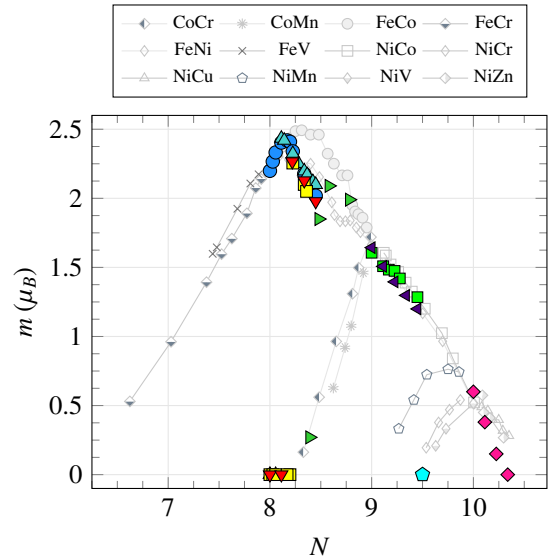

Finally, we want to discuss our results for the magnetic moments shown in Fig. \ref{fig2}  for metallic hydrides in relationship 
to the SP curve. If we superpose our results,  as explained below,  with  the SP curve, we obtain the curve shown in 
Fig. \ref{fig3}. This figure  shows the $m$ for MH$_{x}$ as a function of the effective valence number per metal atom ($N$). 
We used a $d$-band rigid model that has been used to explain the changes of $m$ caused by H \cite{1Exp, Abook}.  
In this model, $N = N_M+\eta x$, where N$_M$ is the number of valence electrons of the metallic host (Fe = 8, Co = 9 
and Ni = 10), and $\eta$ is the number of electrons that are transferred from H to the metal; these approximate values can be obtained 
from our \textit{Ab Initio} calculations ($\eta = 1-Q_{H}$, where $Q_{H}$ is the electronic charge of H inside 
a sphere with the Wigner Seitz radius at equilibrium volume). We found that $\eta$ $\approx$ 0.45 for Fe, Co and Ni, for 
$\alpha$, $\gamma$, $\varepsilon$, $\varepsilon'$ structures, and $\eta$ = 0.39 and 0.5 for Fe ($I4/mmm$ and 
$Pm$-$\bar{3}$$m$, respectively). These values are approximately half of the value of $\eta$ for a substitutional impurity, 
where  $\eta$ = 1. 

As was previously mentioned, in the FeH$_x$ system, it is possible to distinguish both an increasing and a decreasing behavior 
of the $m$,  obtaining a maximum value when $N$ is about 8.2 electrons. Indeed, the elements that are located in the 
ascending part of the SP curve, such as Fe,  have less than half-filled spin  bands.  Therefore, the behavior of their $m$ 
requires a detailed analysis because the majority spin band is not full. Here, the Hund rules play a crucial  role in the 
solid, the pairing energy (spin-up, spin-down pair) and the crystalline field splitting; all have to be computed in order to 
correctly fill the electronic levels.  

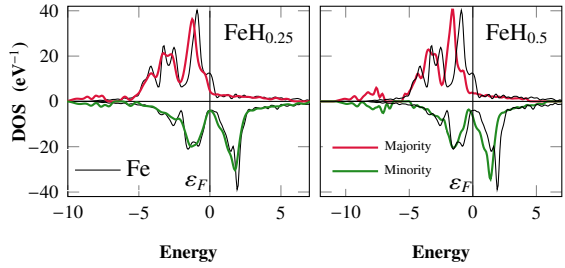
\begin{figure}
 \centering
 \definecolor{color1}{RGB}{30,144,255}
\definecolor{gris0}{RGB}{240,240,240}
\definecolor{gris1}{RGB}{230,230,230}
\definecolor{gris2}{RGB}{200,200,200}
\definecolor{gris3}{RGB}{112,128,144}
\definecolor{gris4}{RGB}{120,120,120}
\definecolor{gris5}{RGB}{255,255,240}

\definecolor{hcp}{RGB}{255,255, 0}
\definecolor{green2}{RGB}{20,235, 34}
\definecolor{green3}{RGB}{34,139, 34}
\definecolor{red2}{RGB}{220,20,60}

\begin{tikzpicture}
		\begin{axis}[
		width=0.61\columnwidth,
		ymin=-42,
		ymax=42,
		xmin=-10,
		xmax=7,
xlabel=\textbf{\scriptsize{Energy}},
ylabel=\textbf{\scriptsize{DOS \hspace{0.04cm} (eV$^{-1}$)}},
		x tick label style={font=\scriptsize},
		y tick label style={font=\scriptsize},
		legend style={font=\small, at={(0.4,0.25)},fill=none ,draw=white},
		every axis y label/.style={at={(ticklabel cs:0.55)},rotate=90,anchor=center}
		]
		\addplot[black] coordinates{(-13,0) (10,0)};
		\addplot[black] coordinates{(0,-40) (0,40)};
		\addplot[black, width=0.2pt] table[x=x,y=y] {dos_h0bcc.dat};
                 \addplot[black, width=0.2pt] table[x=x,y=y1] {dos_h0bcc.dat};
		\addplot[red2, line width=0.8pt] table[x=x,y=y] {dos_h4bcc.dat};
		\addplot[green3,line width=0.8pt] table[x=x,y=y1] {dos_h4bcc.dat};
		 \node at (axis cs: 3.5,30) {\small{FeH$_{0.25}$}};
		 \node at (axis cs: -0.9,-36) {\small{$\varepsilon_{F}$}};
                 \legend{Fe}
		\end{axis}
		\end{tikzpicture}
\hspace{-0.28cm}
\begin{tikzpicture}
		\begin{axis}[
		width=0.61\columnwidth,
		ymin=-42,
		ymax=42,
		xmin=-12,
		xmax=7,
xlabel=\textbf{\scriptsize{Energy}},
		ylabel={},
		x tick label style={font=\scriptsize},
		yticklabels ={},	
		legend style={font=\tiny, at={(0.52,0.35)},draw=white}
		]
		\addplot[red2, line width=0.8pt] table[x=x,y=y] {dos_feh8bcc.dat};
		\addplot[green3,line width=0.8pt] table[x=x,y=y1] {dos_feh8bcc.dat};
		\addplot[black] coordinates{(-13,0) (10,0)};
		\addplot[black] coordinates{(0,-42) (0,42)};
		\addplot[black, width=0.2pt] table[x=x,y=y] {dos_h0bcc.dat};
                 \addplot[black, width=0.2pt] table[x=x,y=y1] {dos_h0bcc.dat};
                 \node at (axis cs: -0.9,-37) {\small{$\varepsilon_{F}$}};
                 \node at (axis cs: 3.5,30) {\small{FeH$_{0.5}$}};
                \legend{Majority,Minority};
		\end{axis}
		\end{tikzpicture}

		
 \caption{Spin dependent DOS of pure Fe (black lines) and FeH$_{x}$ ($x$=0.25 (a) and $x$=0.5 (b)). For Fe and 
 FeH$_{0.25}$, the  Fermi level lies in a  minimum of the minority spin DOS and for FeH$_{0.5}$, the minority spin 
 DOS (green line) is partially filled beyond their minimum.}
\label{fig1}
\end{figure}

From the FeH$_x$ calculations in the $\alpha$ structure in the region where $m$ increases, 
we can see that the Fermi level lies near the minimum of the minority spin DOS, see Fig. \ref{fig1}(a) (pure Fe and 
FeH$_{0.25}$). This result is an indication of the system's structural stability (similar to a filled electronic shell in an atom). 
In these cases, the filling of the bands occurs first in the majority spin bands  (as shown in Fig. \ref{fig1}(a) for 
FeH$_{0.25}$), thus increasing the $m$ of the system, as shown in Fig. \ref{fig2} and also in Fig. \ref{fig3}  
(for $N$ = 8 to $N$ = 8.16 blue circle symbols). After a certain H concentration, the minority spin band begins to be filled 
(see Fig. \ref{fig1}(b)), and the $m$ begins to decrease. Nevertheless, the ferromagnetic materials located in the descending 
part of the SP curve, such as Co and Ni alloys, diminish their magnetic moment as the H concentration increases due to 
the filling of the minority band because the majority 3d band is already filled with 5 electrons.  

Thus, in the case of $\alpha$-FeH$_x$, $m$ increases with $N$ until reaching its maximum value   
($m$ = 2.42  $\mu_{B}$)  with $N$ = 8.16 electrons (i.e. $x$ = 0.375), see Fig. \ref{fig3}, up to the maximum of the SP curve. 
This value is comparable to $m$ = 2.45  $\mu_{B}$  obtained in the FeCo alloy with $N$ = 8.3 electrons. 
When $N$ $>$ 8.16, the $m$ begins to diminish until reaching the value 2.1 $\mu_{B}$ with $N$ = 8.44 valence electrons ($x$ = 1). 
The $m$ diminishes because the majority spin band is filled at $x$ $\simeq$ 0.375; therefore, the 
minority spin band begins to be filled at higher $x$ values, see Fig. \ref{fig1}(b) ($x$ = 0.5). As a consequence, $m$ diminishes, as shown 
in Fig. \ref{fig3}. We also observe that this trend is independent of the structural phase of FeH$_{x}$. 

However, for higher 
concentrations of H at $x$ = 3 ($N$ = 9.5, phase \textit{Pm-$\bar{3}$m}), the tendency of the decreasing behavior of $m$ 
abruptly changes towards the paramagnetic state with $N$ = 9.5 (see the cyan pentagon symbol in Fig. \ref{fig3}).

In the cases of CoH$_x$ and NiH$_x$, their majority spin bands are filled, and only the minority spin band is modified. 
With increasing $N$, the minority spin band is progressively being filled with an effective number of electrons from the 
H atoms. Therefore, $m$ will always diminish independent of the concentration, $x$, and of the structural phase of the system. 

\section{Conclusions}
We have shown that the magnetization $m$ per metal atom of hydrides Fe, Co and Ni as a function of $x$ follows the same trend 
as the Slater-Pauling curve. The magnetic behavior of the studied systems can be understood by considering two main 
effects. The volume expansion of the system caused by the inserted H atoms tends to localize the electronic states, thus 
enhancing ferromagnetism. On the other hand, because of the charge transfer effect, the extra electrons added to the system 
due to H fill the minority spin band for Co and Ni for all the concentrations of H, thus decreasing the magnetic moment. 
Nevertheless, for $\alpha$-Fe, the band filling occurs first in the majority spin band for low concentration of H ($x$ $\leq$ 0.375), 
leading to an increase of the magnetic moment. For $x$ $>$ 0.375, the band filling occurs in the minority spin band causing 
$m$ to decrease, thus following the same rules for the band filling of the ferromagnetic metal alloys in the 
ascending and descending parts of the SP curve.

As a result of these two effects, the overall trend for the magnetic moment per metal atom as a function on the effective 
number of valence electrons follows the SP curve. The presented results and mechanisms for explaining them offer valuable 
insights for the study of magnetism in hydrated binary and tertiary transition metal alloys.

\section*{Acknowledgments}
This work was supported by several projects: the Fondecyt grants 1130950,  3150525, and 1130672; 
the DGIIP-USM grants PI-M-17-3 and 11.15.73-2015; the DGIIP-USM-PIIC project, and the Universidad de San 
Buenaventura grants 9511501-01 and 9511501-02. The authors acknowledge funding from: Financiamiento 
basal para centros cient\'ificos y tecnol\'ogicos de excelencia FB 0807.


\bibliographystyle{plainurl}
\bibliography{ms}

\end{document}